\documentclass[aps,pra,superscriptaddress,twocolumn,showpacs,amsmath,amssymb]{revtex4-1}
\usepackage{bibentry,natbib}
\usepackage{latexsym,bm}
\usepackage{graphicx}
\usepackage{rotating}
\usepackage{amsmath}
\usepackage{rotating}
\usepackage{color}
\usepackage[colorlinks,citecolor=blue]{hyperref}
\usepackage{multirow}
\usepackage{threeparttable}
\usepackage{longtable}

\begin{document}

\title{Complex coordinate rotation method based on gradient optimization}

\author{Zhi-Da Bai}
\affiliation{State Key Laboratory of Magnetic Resonance and Atomic and Molecular Physics,
Wuhan Institute of Physics and Mathematics,
Innovation Academy for Precision Measurement Science and Technology,
Chinese Academy of Sciences, Wuhan 430071, China}
\affiliation{University of Chinese Academy of Sciences, Beijing 100049, China}
\author{Zhen-Xiang Zhong}
\email{zxzhong@wipm.ac.cn}
\affiliation{State Key Laboratory of Magnetic Resonance and Atomic and Molecular Physics,
Wuhan Institute of Physics and Mathematics,
Innovation Academy for Precision Measurement Science and Technology,
Chinese Academy of Sciences, Wuhan 430071, China}
\author{Zong-Chao Yan}
\affiliation{Department of Physics, University of New Brunswick, Fredericton, New Brunswick, Canada E3B 5A3}
\affiliation{State Key Laboratory of Magnetic Resonance and Atomic and Molecular Physics,
Wuhan Institute of Physics and Mathematics,
Innovation Academy for Precision Measurement Science and Technology,
Chinese Academy of Sciences, Wuhan 430071, China}
\affiliation{Center for Cold Atom Physics, Chinese Academy of Sciences, Wuhan 430071, China}
\author{Ting-Yun Shi}
\affiliation{State Key Laboratory of Magnetic Resonance and Atomic and Molecular Physics,
Wuhan Institute of Physics and Mathematics,
Innovation Academy for Precision Measurement Science and Technology,
Chinese Academy of Sciences, Wuhan 430071, China}\date{\today}

\begin{abstract}
   In atomic, molecular, and nuclear physics, the method of complex coordinate rotation is a widely used theoretical tool for studying resonant states. Here, we propose a novel implementation of this method based on the gradient optimization (CCR-GO).
   The main strength of the CCR-GO method is that it does not require manual adjustment of optimization parameters in the wave function; instead,
   a mathematically well-defined optimization path can be followed.
   Our method is proven to be very efficient in searching resonant positions and widths over a variety of few-body atomic systems, and can significantly improve the accuracy of the results. As a special case, the CCR-GO method is equally capable of dealing with bound-state problems with high accuracy, which is traditionally achieved through the usual extreme conditions of energy itself.
\end{abstract}
\maketitle

Resonant states play an important role in atomic, molecular, and nuclear physics and have a long history of research, such as doubly-excited states in two-electron systems~\cite{michishio2011photodetachment,ceeh2011precision},
Efimov states in weakly bound few-body systems~\cite{huang2014observation,zhao2019universal},
resonance phenomena under Debye plasma environment~\cite{ning2015investigation},
four-body resonant states in positronium hydride~\cite{yan2008high} and positron-helium~\cite{yan2018triply},
resonances in positron scattering by atoms and molecules~\cite{sullivan2001search}, metastable states in antiprotonic helium $\bar{\rm p}^{4}\rm He^{+}$~\cite{Korobov1999,korobov2003metastable,korobov2014bethe}, and resonance phenomena in
nuclear physics~\cite{ponomarev1990muon,PhysRevLett.109.072501}.

There exist many theoretical methods for studying resonant states.
In early years, the $S$-matrix~\cite{wheeler1937mathematical} and $R$-matrix~\cite{wigner1947higher} theories were used to solve resonant problems.
In 1970s, the method of complex coordinate rotation (CCR) was mathematically
established~\cite{simon1973resonances}, and was first used in studies of scattering involving
three charged particles by Raju and Doolen~\cite{raju1974scattering}.
After that, the CCR method was further developed to calculate atomic resonant states by Ho~\cite{ho1983method}.
From then on, the CCR method has been widely adopted as a powerful tool for investigating resonant states in atoms and molecules, including its application to high-precision antiprotonic helium 
spectrum~\cite{korobov2003metastable,korobov2014bethe}.
On the other hand, Feshbach in 1962~\cite{feshbach1962unified} formulated a general theory for studying resonances, where the wave function space
is partitioned into closed- and open-channel segments.
The hyperspherical close-coupling method, developed by Lin~\cite{lin1984classification} in 1984 to calculate doubly-excited states, was applied to positron-atom scattering~\cite{igarashi2004stable}.
Recently, the stabilization method combined with hyperspherical coordinates and $B$-spline expansion was applied to positron-atom scattering by Han and co-workers~\cite{han2008ss}.
Among these methods, both the CCR method and the closed-channel approximation of the Feshbach theory can reach high precision for long-lived metastable states with small widths, such as $10^{-11}$ atomic units in $\bar{\rm p}^{4}\rm He^{+}$ decaying via a radiative channel~\cite{Korobov1999}. However, an Auger-dominated state in $\bar{\rm p}^{4}\rm He^{+}$ is usually short-lived and possesses a width larger than $10^{-10}$ atomic units~\cite{Korobov1999}, such as the ($N=31$, $L=30$) state, where $N$ and $L$ are, respectively, the principal and total angular momentum quantum numbers.
The accuracy of the closed-channel approximation of the Feshbach theory is limited by the width of a resonant state~\cite{hu2016variational}, whereas the accuracy of the CCR method can go beyond this limit~\cite{korobov2003metastable}, making the CCR method more suitable for short-lived metastable states.

Since the variational approach using Hylleraas- or Sturmian-type basis sets has been proven to be effective in dealing with atomic or molecular few-body systems, it is natural to combine these basis sets with the CCR
method ~\cite{doolen1975procedure,burgers2000doubly,li2005s,kar2012shape}, and solve resonance problems variationally. However, due to the lack of extreme theorem for a resonance state, historically it is common practice in using the CCR method that the nonlinear variational parameters in the trial wave function are optimized through repeated trial and error manual adjustment, which could become extremely laborious and inefficient, especially for a high-dimensional parameter space.
In this Letter, we propose a novel approach of complex coordinate rotation based on the gradient optimization (CCR-GO). The advantage of the CCR-GO method over the existing resonance methods is that it does not require manual adjustment of nonlinear parameters in the wave function; instead, a mathematically well-defined optimization path can be followed, leading to a resonance pole quickly. Our method will be tested for various three-body atomic systems.

In the method of complex coordinate rotation~\cite{ho1983method}, under the radial coordinate transformation $r\rightarrow r\exp(i\theta)$, the original Hamiltonian of the system $\hat{H}=\hat{T}+\hat{V}$, where $\hat{T}$ and $\hat{V}$ are, respectively, the kinetic and potential energy operators, is transformed into
  \begin{equation}
  \label{ham1}
  \hat{H}\rightarrow \hat{H}(\theta)=\hat{T}\exp(-2i\theta)+\hat{V}\exp(-i\theta)\,,
  \end{equation}
where the rotational angle $\theta$ is assumed to be real and positive.
According to the Balslev-Combes theorem~\cite{simon1973resonances}, in the complex energy plane, for sufficiently large $\theta$ this transformation rotates the continuum spectrum of $\hat{H}$ to ``expose" the resonant poles around the thresholds from the unphysical sheet to physical sheet of the Riemann surface, and the bound state poles remain unchanged on the negative side of the real axis. The eigenenergies can be obtained by solving the following complex eigenvalue problem
\begin{equation}
  \label{problem1}
  \hat{H}(\theta)\Psi_{\theta}=E\Psi_{\theta}\,,
\end{equation}
where the eigenfunction $\Psi_{\theta}$ is square integrable and the corresponding discrete complex eigenvalue $E=E_{r}-i\Gamma/2$ defines the position $E_{r}$ and the width $\Gamma$ of a resonance.
By choosing a basis set $\{\psi_n,n=1,\ldots, \mathcal{N}\}$ in an $\mathcal{N}$-dimensional Hilbert space, the complex eigenvalue problem~(\ref{problem1}) can be converted to the following generalized algebraic complex eigenvalue problem
\begin{equation}
\label{problem2}
{\bf H}(\theta)\Psi_{\theta}=E{\bf O}\Psi_{\theta}\,,
\end{equation}
where ${\bf H}(\theta)_{ij}=\langle\psi_{i}|\hat{H}(\theta)|\psi_{j}\rangle$ are the $\mathcal{N}\times \mathcal{N}$ Hamiltonian matrix elements and ${\bf O}_{ij}=\langle\psi_{i}|\psi_{j}\rangle$ are the overlap matrix elements.
Since a resonance wave function is square integrable, the rotated Hamiltonian $\hat{H}(\theta)$ holds the complex variational principle that
makes the complex energy eigenvalue stationary, although not necessarily extreme, with respect to any parameter $\xi$ in the wave function, such as the rotational angle $\theta$, or a nonlinear parameter in a Hylleraas basis set, or the box size of a $B$-spline basis set, {\it i.e.},
\begin{equation}
\label{rescond}
\partial_\xi E\equiv \frac{\partial E}{\partial \xi}=0
\end{equation}
at a resonance pole.
This expression can be understood as a stability condition for a resonant energy, which of course also applies to any bound state as a special case.
Since we do not have the extreme theorem for a resonance energy $E$ in general, instead of dealing with $E$ itself, we focus on
$|\partial_\xi E|$ and minimize it by varying $\xi$, due to the obvious fact that $|\partial_\xi E|\ge 0$. This is the essence of our CCR-GO method.

\begin{figure}[h]
	\centering
	\includegraphics[scale=0.3]{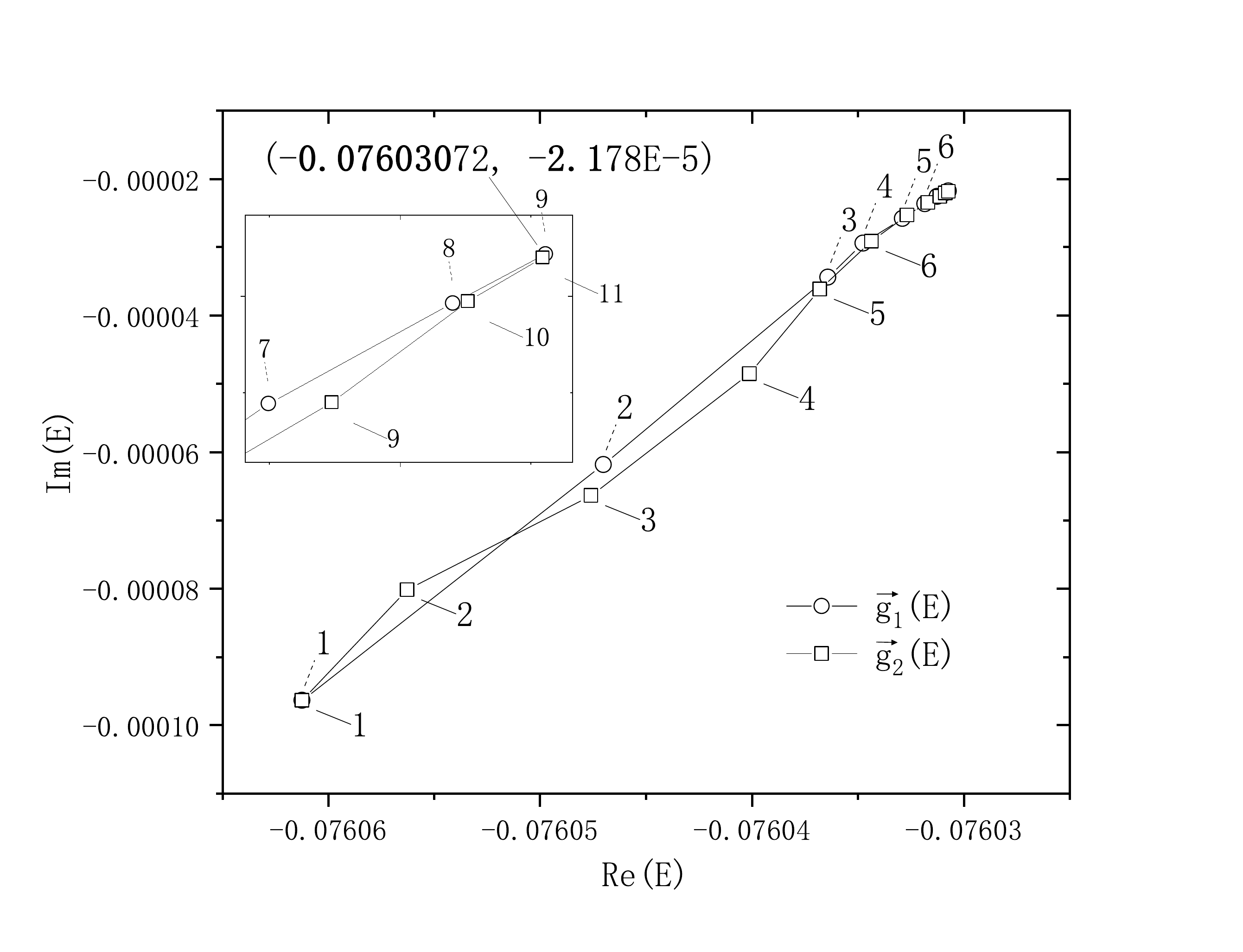}
	\caption{Two optimization paths based on $\vec{g}_1(E)$ and $\vec{g}_2(E)$ for the lowest resonant state $^1\!S^{\rm e}$ in $\textrm{Ps}^{-}$ below the Ps~($N=2$) threshold, with the size of basis set $\mathcal{N}=252$.
		The inset is an enlarged view of the paths around the convergence point. In atomic units.}\label{convergefig}
\end{figure}

\begin{figure}[h]
	\centering
	\includegraphics[scale=0.3]{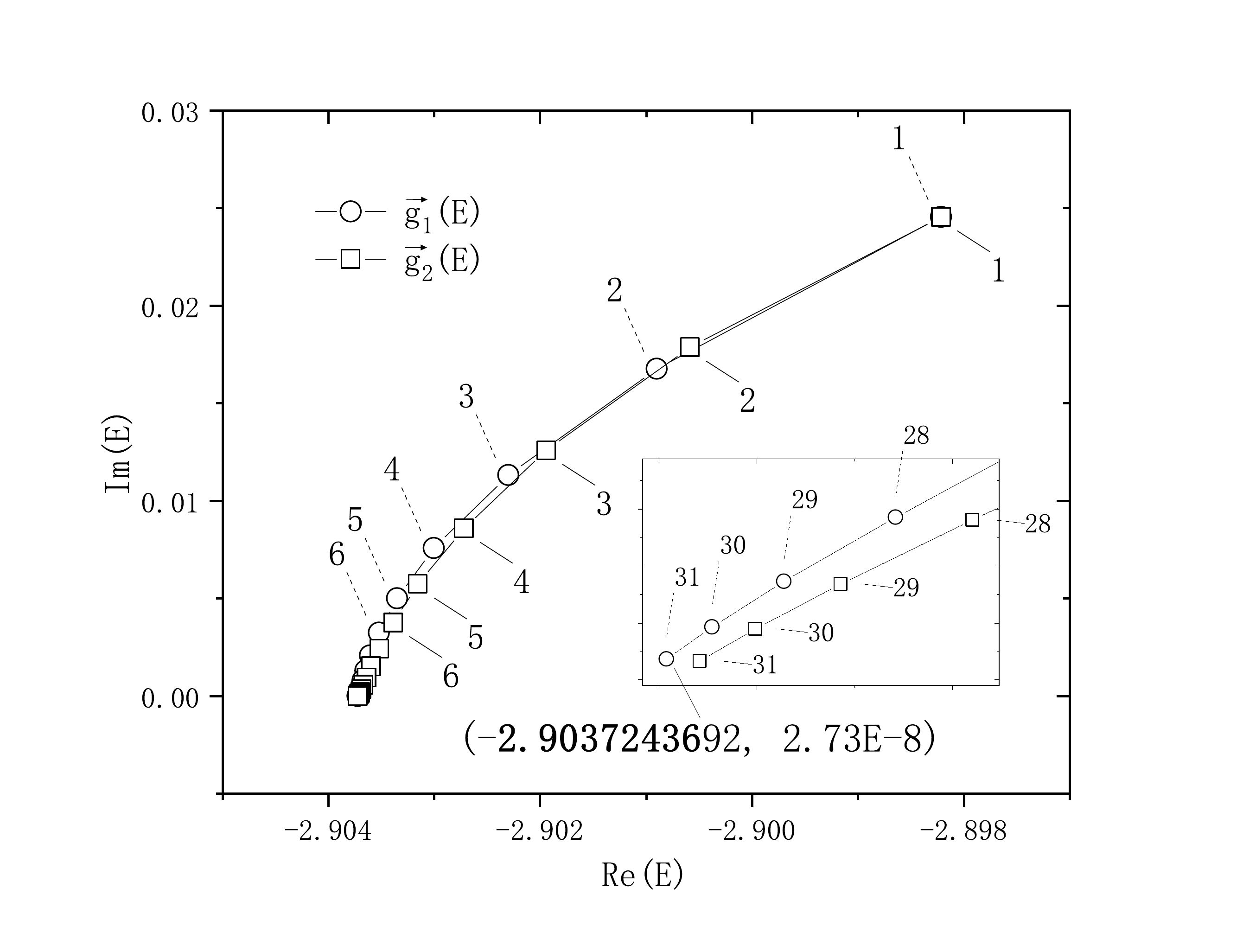}
	\caption{Two optimization paths based on $\vec{g}_1(E)$ and $\vec{g}_2(E)$  for the ground state of helium, with the size of basis set $\mathcal{N}=252$. The inset is an enlarged view of the paths around the convergence point. In atomic units.}\label{boundfig1}
\end{figure}

\begin{table*}[ht]
\begin{threeparttable}
\caption{Resonance parameters $(E_r,\Gamma/2)$ for various three-body Coulombic systems.
In the table, ${\mathcal N}$ is the size of basis set.
A comparison with some of the best theoretical results is also presented. In atomic units.\label{comparision1}}
\begin{tabular}{lccll}
\hline
\hline
Author (year)&Ref.&${\mathcal N}$&\multicolumn{1}{c}{$-E_r$}&\multicolumn{1}{c}{$\Gamma/2$}  \\
\hline
\multicolumn{5}{c}{$\textrm{Ps}^{-}$ $^{1}\!{S}^{\rm e}$, below Ps~($N=2$) threshold}\\
Ho (1979)&\cite{ho1979autoionization}&161        & 0.076030(1)& 0.000021(1)\\
Li and Shakeshaft (2005)&\cite{li2005s} &10206 &0.07603044235&0.00002151725\\
This work\tnote{1}&&1222& 0.07603044186(2)& 0.00002151695(1)\\
\multicolumn{5}{c}{$\textrm{H}^{-}$ $S$-wave shape resonance, above H~($N=2$) threshold}\\
B$\ddot{\text{u}}$rgers and Lindroth (2000)&\cite{burgers2000doubly}&34447& 0.103035676  &0.015627312 \\
Kar and Ho (2012)&\cite{kar2012shape}&700  & 0.1030357(50) &0.0156273(50)\\
This work\tnote{1}&                  &1222 & 0.103035677(3)&0.015627312(3)\\
\multicolumn{5}{c}{He $^{1}\!{S}^{\rm e}$(1), below H~($N=2$) threshold}\\
Ho (1981)&\cite{ho1981complex}&161 &0.77787&	0.00227 \\
Gning {\it et al.} (2015)&\cite{GNING20151}&& 0.777865	&0.002265  \\
This work\tnote{2} &&1925&0.7778675(3)&0.0022706(2)\\
This work\tnote{1} &&715& 0.7778676356(3)&0.0022706527(1)\\
\multicolumn{5}{c}{He $^{3}\!{P}^{\rm o}$(1), below H~($N=2$) threshold}\\
Ho (1981)&\cite{ho1981complex}&165& 0.7604975    &	0.0001485          \\
This work\tnote{1} &&969& 0.76049238762(3)&	0.0001494308(1)  \\
\multicolumn{5}{c}{He $^{1}\!D^{\rm e}$(1), below H~($N=2$) threshold}\\
Ho and Bhatia (1991)&\cite{ho1991complex}&1230& 0.7019457&	0.0011811        \\
This work\tnote{1} &&959&  0.70194550(1)&	0.001181226(3)        \\
\multicolumn{5}{c}{$\bar{\rm p}^{4}\rm He^{+}$ $(N=31,L=30)$}\\
Korobov (2014) &\cite{korobov2014bethe}&7000&3.67977478748142(4)&4.76010$\times10^{-9}$\\
 This work\tnote{1}   &&2555& 3.67977478748(1) & 4.754(2)$\times10^{-9}$\\
\hline
\hline
\end{tabular}
\begin{tablenotes}
\item[1] Basis set Eq.~(\ref{complexfunc}) with complex nonlinear parameters.
\item[2] Basis set Eq.~(\ref{realfunc}) with real nonlinear parameters.
\end{tablenotes}
\end{threeparttable}
\end{table*}

To be specific, let us consider a three-body Coulombic system, such as Ps$^-$, H$^-$, He, and $\bar{\rm p}^{4}\rm He^{+}$. After eliminating the center of mass coordinates, a three-body problem is reduced to a quasi two-body one with $\vec{r}_1$ and $\vec{r}_2$ being their position vectors relative to the third particle.
In order to solve the complex eigenvalue problem~(\ref{problem1}), we use two types of basis sets.
The first one consists of Hylleraas functions with real nonlinear parameters $\alpha$ and $\beta$:
\begin{equation}
\label{realfunc}
\{r^{\ell}_{1}r^{m}_{2}r^{n}_{12}e^{-\alpha r_{1}-\beta r_{2}}\mathcal{Y}^{LM}_{\ell_1\ell_2}(\hat{r}_1,\hat{r}_2)\}\,,
\end{equation}
where $\mathcal{Y}^{LM}_{\ell_1\ell_2}(\hat{r}_1,\hat{r}_2)$ is the angular momenta $(\ell_1,\ell_2)$-coupled spherical harmonics to form a common eigenstate of $L^2$ and $L_z$.
It is noted that a proper symmetrization of the final wave function is implied for a system containing two identical particles. The possible values of $\ell_1$ and $\ell_2$ are those fulfilling $\ell_{1}+\ell_{2}=L$ for a state of natural parity $(-1)^{L}$ or $\ell_{1}+\ell_{2}=L+1$ for a state of unnatural parity $(-1)^{L+1}$.
Each configuration $(\ell_{1},\ell_{2})$ has its own set of nonlinear parameters. In order to enhance the rate of convergence, we may further divide the most important configuration into more sub-groups each having different set of nonlinear parameters.
The basis set is generated by including all terms such that ${\ell}+m+n\leq\Omega$ with integer $\Omega$ controlling the size of basis set.
More information about the construction of basis sets can be found in~\cite{Drake1988,hu2016variational}.
This type of basis set has been widely applied to three-body atomic and molecular systems, such as helium~\cite{Drake1988}, the hydrogen molecular ions~\cite{yan2003energies}, and antiprotonic helium~\cite{hu2016variational}.
The second type of basis set consists of Hylleraas basis functions containing complex nonlinear parameters:
\begin{equation}
\label{complexfunc}
\{r^{\ell}_{1}r^{m}_{2}r^{n}_{12}e^{-(\alpha+i\mu)r_{1}-(\beta+i\nu)r_{2}-
(\gamma+i\sigma)r_{12}}\mathcal{Y}^{LM}_{\ell_{1}\ell_{2}}(\hat{r}_1,\hat{r}_2)\}\,,
\end{equation}
where $\alpha$, $\beta$, $\gamma$, $\mu$, $\nu$, and $\sigma$ are real. It is noted that, with nonzero imaginary parts of nonlinear parameters in the basis set, the wave function is more capable of describing the asymptotic behavior of a resonance state;
it can also be more flexible to reflect molecular characteristics for some exotic systems, such as
$\bar{\rm p}^{4}\rm He^{+}$~\cite{korobov2014bethe}.

In our CCR-GO approach, we try to minimize either
\begin{equation}
\label{optfunc1}
f_{1}(\boldsymbol{\mathcal{P}})=|\partial_{\theta} E|
\end{equation}
or
\begin{eqnarray}
\label{optfunc2}
f_{2}(\boldsymbol{\mathcal{P}})&=&|\partial_{\theta} E|+|\partial_{\alpha} E|+|\partial_{\beta} E|+|\partial_{\gamma} E|\nonumber\\
&+&|\partial_{\mu} E|+|\partial_{\nu} E|+|\partial_{\sigma} E|\,,
\end{eqnarray}
where $\boldsymbol{\mathcal{P}} \equiv (\theta,\alpha,\beta,\gamma,\mu,\nu,\sigma)$ aggregates all variational parameters, such as the rotational angle $\theta$ and the nonlinear parameters $\alpha$, $\beta$, $\gamma$, $\mu$, $\nu$, and $\sigma$ that appear in Eq.~(\ref{complexfunc}), for example. The optimization procedure for minimizing $f_1(\boldsymbol{\mathcal{P}})$ or $f_2(\boldsymbol{\mathcal{P}})$ can be done iteratively
from step $k$ to step $k+1$: $[\boldsymbol{\mathcal{P}}]_k\rightarrow [\boldsymbol{\mathcal{P}}]_{k+1}$,
where the initial values of optimization parameters, as well as the search directions and steps for each parameters, are determined by the Broyden-Fletcher-Goldfarb-Shanno algorithm~\cite{fletcher2013practical}, or the quasi-Newton method. In particular, the search directions are directly related to the following vectors
\begin{equation}
\label{opttheta}
\vec{g}_{1}(E)=\nabla f_{1}(\boldsymbol{\mathcal{P}})\,,\ \
\vec{g}_{2}(E)=\nabla f_{2}(\boldsymbol{\mathcal{P}})\,,
\end{equation}
where $\nabla=(\partial_{\theta},\partial_{\alpha},\partial_{\beta},\partial_{\gamma},\partial_{\mu},\partial_{\nu},\partial_{\sigma})$ is the gradient operator in the full parameter space. It is noted here that, for an efficient search of a resonant state, the rotational angle term $|\partial_{\theta} E|$ must be included in the optimization procedure, whereas the other parameters can sometimes be optional except for some broad resonant states.

For the purpose of demonstrating the effectiveness of our optimization technique, Fig.~\ref{convergefig} shows the convergence pattern for
the lowest $S$-wave resonant state in Ps$^-$ using only one set of nonlinear parameters in Eq.~(\ref{complexfunc}), with the size of basis set $\mathcal{N}=252$.
One can see from the figure that, starting from the same initial point, the two optimization paths based on $\vec{g}_1(E)$ and $\vec{g}_2(E)$
approach each other rapidly around the resonant pole after about 9 iterations. Of course, further iterations are needed if one wishes to obtain higher accuracy, as shown in Table~\ref{comparision1}.
It is noted that, since the resonance width of this state is relatively small, the searching process is less sensitive to the initial values of nonlinear parameters and the searching paths. However, for a resonant state of broad width, it is advisable to use the more demanding condition $\vec{g}_2(E)$ to do searching, together with suitable initial values of nonlinear parameters.
It is also noted that our CCR-GO method is applicable not only to resonant states, but also to bound ones.
Figure~\ref{boundfig1} shows two optimization paths determined by $\vec{g}_1(E)$ and $\vec{g}_2(E)$ for locating the ground state of helium, with the size of basis set $\mathcal{N}=252$. The ground-state energy, thus obtained after 30 iterations, is accurate to about 9 digits.

Table~\ref{comparision1} lists the resonance positions and widths using the CCR-GO method for some representative Coulombic systems, including the weakly-bound Ps$^-$ and H$^-$, the tightly-bound He, and the exotic quasi-molecule $\bar{\rm p}^{4}\rm He^{+}$, with the angular momentum
quantum number $L$ ranging from $0$ to $30$ and the resonance width ranging from $10^{-2}$ to $10^{-9}$ atomic units.
Also in the table, a comparison is made with some of the best calculations in the literature. One can see from the table that
with moderately large sizes of basis sets, our method can not only reproduce but also be capable of significantly improving the previous values for Ps$^-$, H$^-$, and He. To our knowledge, our result for $\bar{\rm p}^{4}\rm He^{+}$ is the only theoretical value that confirms the Korobov's calculation for both the position and width, although much smaller size of basis set is used in our work.
Our calculations show that the Hylleraas basis sets with complex nonlinear parameters are more powerful than those with real ones in achieving higher precision.

To sum up, we have presented a new approach called the CCR-GO method, which for the first time makes the search for resonance mathematically automated. This is in sharp contrast to the traditional way of manual adjustment of variational parameters. Therefore, our method can greatly improve the search efficiency and search accuracy of resonance poles. Resonance phenomena exist ubiquitously in physics.
The significance of our method is by no means limited to few-body atomic systems; it can also be applied in principle to find resonance poles in many areas of physics, including nuclear and elementary particle physics.

This work was supported by the National Natural Science Foundation of China with Grants Nos. 91636216, 11974382, and 11474316, by the Chinese Academy of Sciences Strategic Priority Research Program with Grant No. XDB21020200, and by the YIPA program. ZXZ would like to thank V. I. Korobov of JINR for providing his computer codes of complex nonlinear parameters. ZCY
acknowledges the support of NSERC, SHARCnet, and ACEnet of Canada.

\bibliography{ttlib}

\begin{thebibliography}{34}%
\makeatletter
\providecommand \@ifxundefined [1]{%
 \@ifx{#1\undefined}
}%
\providecommand \@ifnum [1]{%
 \ifnum #1\expandafter \@firstoftwo
 \else \expandafter \@secondoftwo
 \fi
}%
\providecommand \@ifx [1]{%
 \ifx #1\expandafter \@firstoftwo
 \else \expandafter \@secondoftwo
 \fi
}%
\providecommand \natexlab [1]{#1}%
\providecommand \enquote  [1]{``#1''}%
\providecommand \bibnamefont  [1]{#1}%
\providecommand \bibfnamefont [1]{#1}%
\providecommand \citenamefont [1]{#1}%
\providecommand \href@noop [0]{\@secondoftwo}%
\providecommand \href [0]{\begingroup \@sanitize@url \@href}%
\providecommand \@href[1]{\@@startlink{#1}\@@href}%
\providecommand \@@href[1]{\endgroup#1\@@endlink}%
\providecommand \@sanitize@url [0]{\catcode `\\12\catcode `\$12\catcode
  `\&12\catcode `\#12\catcode `\^12\catcode `\_12\catcode `\%12\relax}%
\providecommand \@@startlink[1]{}%
\providecommand \@@endlink[0]{}%
\providecommand \url  [0]{\begingroup\@sanitize@url \@url }%
\providecommand \@url [1]{\endgroup\@href {#1}{\urlprefix }}%
\providecommand \urlprefix  [0]{URL }%
\providecommand \Eprint [0]{\href }%
\providecommand \doibase [0]{http://dx.doi.org/}%
\providecommand \selectlanguage [0]{\@gobble}%
\providecommand \bibinfo  [0]{\@secondoftwo}%
\providecommand \bibfield  [0]{\@secondoftwo}%
\providecommand \translation [1]{[#1]}%
\providecommand \BibitemOpen [0]{}%
\providecommand \bibitemStop [0]{}%
\providecommand \bibitemNoStop [0]{.\EOS\space}%
\providecommand \EOS [0]{\spacefactor3000\relax}%
\providecommand \BibitemShut  [1]{\csname bibitem#1\endcsname}%
\let\auto@bib@innerbib\@empty
\bibitem [{\citenamefont {Michishio}\ \emph {et~al.}(2011)\citenamefont
  {Michishio}, \citenamefont {Tachibana}, \citenamefont {Terabe}, \citenamefont
  {Igarashi}, \citenamefont {Wada}, \citenamefont {Kuga}, \citenamefont
  {Yagishita}, \citenamefont {Hyodo},\ and\ \citenamefont
  {Nagashima}}]{michishio2011photodetachment}%
  \BibitemOpen
  \bibfield  {author} {\bibinfo {author} {\bibfnamefont {K.}~\bibnamefont
  {Michishio}}, \bibinfo {author} {\bibfnamefont {T.}~\bibnamefont
  {Tachibana}}, \bibinfo {author} {\bibfnamefont {H.}~\bibnamefont {Terabe}},
  \bibinfo {author} {\bibfnamefont {A.}~\bibnamefont {Igarashi}}, \bibinfo
  {author} {\bibfnamefont {K.}~\bibnamefont {Wada}}, \bibinfo {author}
  {\bibfnamefont {T.}~\bibnamefont {Kuga}}, \bibinfo {author} {\bibfnamefont
  {A.}~\bibnamefont {Yagishita}}, \bibinfo {author} {\bibfnamefont
  {T.}~\bibnamefont {Hyodo}}, \ and\ \bibinfo {author} {\bibfnamefont
  {Y.}~\bibnamefont {Nagashima}},\ }\href {\doibase
  10.1103/physrevlett.106.153401} {\bibfield  {journal} {\bibinfo  {journal}
  {Phys. Rev. Lett.}\ }\textbf {\bibinfo {volume} {106}},\ \bibinfo {pages}
  {153401} (\bibinfo {year} {2011})}\BibitemShut {NoStop}%
\bibitem [{\citenamefont {Ceeh}\ \emph {et~al.}(2011)\citenamefont {Ceeh},
  \citenamefont {Hugenschmidt}, \citenamefont {Schreckenbach}, \citenamefont
  {G\"artner}, \citenamefont {Thirolf}, \citenamefont {Fleischer},\ and\
  \citenamefont {Schwalm}}]{ceeh2011precision}%
  \BibitemOpen
  \bibfield  {author} {\bibinfo {author} {\bibfnamefont {H.}~\bibnamefont
  {Ceeh}}, \bibinfo {author} {\bibfnamefont {C.}~\bibnamefont {Hugenschmidt}},
  \bibinfo {author} {\bibfnamefont {K.}~\bibnamefont {Schreckenbach}}, \bibinfo
  {author} {\bibfnamefont {S.~A.}\ \bibnamefont {G\"artner}}, \bibinfo {author}
  {\bibfnamefont {P.~G.}\ \bibnamefont {Thirolf}}, \bibinfo {author}
  {\bibfnamefont {F.}~\bibnamefont {Fleischer}}, \ and\ \bibinfo {author}
  {\bibfnamefont {D.}~\bibnamefont {Schwalm}},\ }\href {\doibase
  10.1103/physreva.84.062508} {\bibfield  {journal} {\bibinfo  {journal} {Phys.
  Rev. A}\ }\textbf {\bibinfo {volume} {84}},\ \bibinfo {pages} {062508}
  (\bibinfo {year} {2011})}\BibitemShut {NoStop}%
\bibitem [{\citenamefont {Huang}\ \emph {et~al.}(2014)\citenamefont {Huang},
  \citenamefont {Sidorenkov}, \citenamefont {Grimm},\ and\ \citenamefont
  {Hutson}}]{huang2014observation}%
  \BibitemOpen
  \bibfield  {author} {\bibinfo {author} {\bibfnamefont {B.}~\bibnamefont
  {Huang}}, \bibinfo {author} {\bibfnamefont {L.~A.}\ \bibnamefont
  {Sidorenkov}}, \bibinfo {author} {\bibfnamefont {R.}~\bibnamefont {Grimm}}, \
  and\ \bibinfo {author} {\bibfnamefont {J.~M.}\ \bibnamefont {Hutson}},\
  }\href {\doibase 10.1103/physrevlett.112.190401} {\bibfield  {journal}
  {\bibinfo  {journal} {Phys. Rev. Lett.}\ }\textbf {\bibinfo {volume} {112}},\
  \bibinfo {pages} {190401} (\bibinfo {year} {2014})}\BibitemShut {NoStop}%
\bibitem [{\citenamefont {Zhao}\ \emph {et~al.}(2019)\citenamefont {Zhao},
  \citenamefont {Han}, \citenamefont {Wu},\ and\ \citenamefont
  {Shi}}]{zhao2019universal}%
  \BibitemOpen
  \bibfield  {author} {\bibinfo {author} {\bibfnamefont {C.-Y.}\ \bibnamefont
  {Zhao}}, \bibinfo {author} {\bibfnamefont {H.-L.}\ \bibnamefont {Han}},
  \bibinfo {author} {\bibfnamefont {M.-S.}\ \bibnamefont {Wu}}, \ and\ \bibinfo
  {author} {\bibfnamefont {T.-Y.}\ \bibnamefont {Shi}},\ }\href {\doibase
  10.1103/physreva.100.052702} {\bibfield  {journal} {\bibinfo  {journal}
  {Phys. Rev. A}\ }\textbf {\bibinfo {volume} {100}},\ \bibinfo {pages}
  {052702} (\bibinfo {year} {2019})}\BibitemShut {NoStop}%
\bibitem [{\citenamefont {Ning}\ \emph {et~al.}(2015)\citenamefont {Ning},
  \citenamefont {Yan},\ and\ \citenamefont {Ho}}]{ning2015investigation}%
  \BibitemOpen
  \bibfield  {author} {\bibinfo {author} {\bibfnamefont {Y.}~\bibnamefont
  {Ning}}, \bibinfo {author} {\bibfnamefont {Z.-C.}\ \bibnamefont {Yan}}, \
  and\ \bibinfo {author} {\bibfnamefont {Y.~K.}\ \bibnamefont {Ho}},\ }\href
  {\doibase 10.1063/1.4906363} {\bibfield  {journal} {\bibinfo  {journal}
  {Phys. Plasmas}\ }\textbf {\bibinfo {volume} {22}},\ \bibinfo {pages}
  {013302} (\bibinfo {year} {2015})}\BibitemShut {NoStop}%
\bibitem [{\citenamefont {Yan}\ and\ \citenamefont {Ho}(2008)}]{yan2008high}%
  \BibitemOpen
  \bibfield  {author} {\bibinfo {author} {\bibfnamefont {Z.-C.}\ \bibnamefont
  {Yan}}\ and\ \bibinfo {author} {\bibfnamefont {Y.~K.}\ \bibnamefont {Ho}},\
  }\href {\doibase 10.1103/physreva.78.012711} {\bibfield  {journal} {\bibinfo
  {journal} {Phys. Rev. A}\ }\textbf {\bibinfo {volume} {78}},\ \bibinfo
  {pages} {012711} (\bibinfo {year} {2008})}\BibitemShut {NoStop}%
\bibitem [{\citenamefont {Yan}\ and\ \citenamefont {Ho}(2018)}]{yan2018triply}%
  \BibitemOpen
  \bibfield  {author} {\bibinfo {author} {\bibfnamefont {Z.-C.}\ \bibnamefont
  {Yan}}\ and\ \bibinfo {author} {\bibfnamefont {Y.~K.}\ \bibnamefont {Ho}},\
  }\href {\doibase 10.1103/physreva.98.062702} {\bibfield  {journal} {\bibinfo
  {journal} {Phys. Rev. A}\ }\textbf {\bibinfo {volume} {98}},\ \bibinfo
  {pages} {062702} (\bibinfo {year} {2018})}\BibitemShut {NoStop}%
\bibitem [{\citenamefont {Sullivan}\ \emph {et~al.}(2001)\citenamefont
  {Sullivan}, \citenamefont {Gilbert}, \citenamefont {Buckman},\ and\
  \citenamefont {Surko}}]{sullivan2001search}%
  \BibitemOpen
  \bibfield  {author} {\bibinfo {author} {\bibfnamefont {J.~P.}\ \bibnamefont
  {Sullivan}}, \bibinfo {author} {\bibfnamefont {S.~J.}\ \bibnamefont
  {Gilbert}}, \bibinfo {author} {\bibfnamefont {S.~J.}\ \bibnamefont
  {Buckman}}, \ and\ \bibinfo {author} {\bibfnamefont {C.~M.}\ \bibnamefont
  {Surko}},\ }\href {\doibase 10.1088/0953-4075/34/15/102} {\bibfield
  {journal} {\bibinfo  {journal} {J. Phys. B}\ }\textbf {\bibinfo {volume}
  {34}},\ \bibinfo {pages} {L467} (\bibinfo {year} {2001})}\BibitemShut
  {NoStop}%
\bibitem [{\citenamefont {Korobov}\ \emph {et~al.}(1999)\citenamefont
  {Korobov}, \citenamefont {Bakalov},\ and\ \citenamefont
  {Monkhorst}}]{Korobov1999}%
  \BibitemOpen
  \bibfield  {author} {\bibinfo {author} {\bibfnamefont {V.~I.}\ \bibnamefont
  {Korobov}}, \bibinfo {author} {\bibfnamefont {D.}~\bibnamefont {Bakalov}}, \
  and\ \bibinfo {author} {\bibfnamefont {H.~J.}\ \bibnamefont {Monkhorst}},\
  }\href {\doibase 10.1103/physreva.59.r919} {\bibfield  {journal} {\bibinfo
  {journal} {Phys. Rev. A}\ }\textbf {\bibinfo {volume} {59}},\ \bibinfo
  {pages} {R919(R)} (\bibinfo {year} {1999})}\BibitemShut {NoStop}%
\bibitem [{\citenamefont {Korobov}(2003)}]{korobov2003metastable}%
  \BibitemOpen
  \bibfield  {author} {\bibinfo {author} {\bibfnamefont {V.~I.}\ \bibnamefont
  {Korobov}},\ }\href {\doibase 10.1103/physreva.67.062501} {\bibfield
  {journal} {\bibinfo  {journal} {Phys. Rev. A}\ }\textbf {\bibinfo {volume}
  {67}},\ \bibinfo {pages} {062501} (\bibinfo {year} {2003})}\BibitemShut
  {NoStop}%
\bibitem [{\citenamefont {Korobov}(2014)}]{korobov2014bethe}%
  \BibitemOpen
  \bibfield  {author} {\bibinfo {author} {\bibfnamefont {V.~I.}\ \bibnamefont
  {Korobov}},\ }\href {\doibase 10.1103/physreva.89.014501} {\bibfield
  {journal} {\bibinfo  {journal} {Phys. Rev. A}\ }\textbf {\bibinfo {volume}
  {89}},\ \bibinfo {pages} {014501} (\bibinfo {year} {2014})}\BibitemShut
  {NoStop}%
\bibitem [{\citenamefont {Ponomarev}(1990)}]{ponomarev1990muon}%
  \BibitemOpen
  \bibfield  {author} {\bibinfo {author} {\bibfnamefont {L.~I.}\ \bibnamefont
  {Ponomarev}},\ }\href {\doibase 10.1080/00107519008222019} {\bibfield
  {journal} {\bibinfo  {journal} {Contemp. Phys.}\ }\textbf {\bibinfo {volume}
  {31}},\ \bibinfo {pages} {219} (\bibinfo {year} {1990})}\BibitemShut
  {NoStop}%
\bibitem [{\citenamefont {Lu}\ \emph {et~al.}(2012)\citenamefont {Lu},
  \citenamefont {Zhao},\ and\ \citenamefont {Zhou}}]{PhysRevLett.109.072501}%
  \BibitemOpen
  \bibfield  {author} {\bibinfo {author} {\bibfnamefont {B.-N.}\ \bibnamefont
  {Lu}}, \bibinfo {author} {\bibfnamefont {E.-G.}\ \bibnamefont {Zhao}}, \ and\
  \bibinfo {author} {\bibfnamefont {S.-G.}\ \bibnamefont {Zhou}},\ }\href
  {\doibase 10.1103/PhysRevLett.109.072501} {\bibfield  {journal} {\bibinfo
  {journal} {Phys. Rev. Lett.}\ }\textbf {\bibinfo {volume} {109}},\ \bibinfo
  {pages} {072501} (\bibinfo {year} {2012})}\BibitemShut {NoStop}%
\bibitem [{\citenamefont {Wheeler}(1937)}]{wheeler1937mathematical}%
  \BibitemOpen
  \bibfield  {author} {\bibinfo {author} {\bibfnamefont {J.~A.}\ \bibnamefont
  {Wheeler}},\ }\href {\doibase 10.1103/physrev.52.1107} {\bibfield  {journal}
  {\bibinfo  {journal} {Phys. Rev.}\ }\textbf {\bibinfo {volume} {52}},\
  \bibinfo {pages} {1107} (\bibinfo {year} {1937})}\BibitemShut {NoStop}%
\bibitem [{\citenamefont {Wigner}\ and\ \citenamefont
  {Eisenbud}(1947)}]{wigner1947higher}%
  \BibitemOpen
  \bibfield  {author} {\bibinfo {author} {\bibfnamefont {E.~P.}\ \bibnamefont
  {Wigner}}\ and\ \bibinfo {author} {\bibfnamefont {L.}~\bibnamefont
  {Eisenbud}},\ }\href {\doibase 10.1103/physrev.72.29} {\bibfield  {journal}
  {\bibinfo  {journal} {Phys. Rev.}\ }\textbf {\bibinfo {volume} {72}},\
  \bibinfo {pages} {29} (\bibinfo {year} {1947})}\BibitemShut {NoStop}%
\bibitem [{\citenamefont {Simon}(1973)}]{simon1973resonances}%
  \BibitemOpen
  \bibfield  {author} {\bibinfo {author} {\bibfnamefont {B.}~\bibnamefont
  {Simon}},\ }\href {\doibase 10.2307/1970847} {\bibfield  {journal} {\bibinfo
  {journal} {Ann. Math.}\ }\textbf {\bibinfo {volume} {97}},\ \bibinfo {pages}
  {247} (\bibinfo {year} {1973})}\BibitemShut {NoStop}%
\bibitem [{\citenamefont {Raju}\ and\ \citenamefont
  {Doolen}(1974)}]{raju1974scattering}%
  \BibitemOpen
  \bibfield  {author} {\bibinfo {author} {\bibfnamefont {S.~B.}\ \bibnamefont
  {Raju}}\ and\ \bibinfo {author} {\bibfnamefont {G.}~\bibnamefont {Doolen}},\
  }\href {\doibase 10.1103/physreva.9.1965} {\bibfield  {journal} {\bibinfo
  {journal} {Phys. Rev. A}\ }\textbf {\bibinfo {volume} {9}},\ \bibinfo {pages}
  {1965} (\bibinfo {year} {1974})}\BibitemShut {NoStop}%
\bibitem [{\citenamefont {Ho}(1983)}]{ho1983method}%
  \BibitemOpen
  \bibfield  {author} {\bibinfo {author} {\bibfnamefont {Y.~K.}\ \bibnamefont
  {Ho}},\ }\href {\doibase 10.1016/0370-1573(83)90112-6} {\bibfield  {journal}
  {\bibinfo  {journal} {Phys. Rep.}\ }\textbf {\bibinfo {volume} {99}},\
  \bibinfo {pages} {1} (\bibinfo {year} {1983})}\BibitemShut {NoStop}%
\bibitem [{\citenamefont {Feshbach}(1962)}]{feshbach1962unified}%
  \BibitemOpen
  \bibfield  {author} {\bibinfo {author} {\bibfnamefont {H.}~\bibnamefont
  {Feshbach}},\ }\href {\doibase 10.1016/0003-4916(62)90221-x} {\bibfield
  {journal} {\bibinfo  {journal} {Ann. Phys.}\ }\textbf {\bibinfo {volume}
  {19}},\ \bibinfo {pages} {287} (\bibinfo {year} {1962})}\BibitemShut
  {NoStop}%
\bibitem [{\citenamefont {Lin}(1984)}]{lin1984classification}%
  \BibitemOpen
  \bibfield  {author} {\bibinfo {author} {\bibfnamefont {C.~D.}\ \bibnamefont
  {Lin}},\ }\href {\doibase 10.1103/physreva.29.1019} {\bibfield  {journal}
  {\bibinfo  {journal} {Phys. Rev. A}\ }\textbf {\bibinfo {volume} {29}},\
  \bibinfo {pages} {1019} (\bibinfo {year} {1984})}\BibitemShut {NoStop}%
\bibitem [{\citenamefont {Igarashi}\ and\ \citenamefont
  {Shimamura}(2004)}]{igarashi2004stable}%
  \BibitemOpen
  \bibfield  {author} {\bibinfo {author} {\bibfnamefont {A.}~\bibnamefont
  {Igarashi}}\ and\ \bibinfo {author} {\bibfnamefont {I.}~\bibnamefont
  {Shimamura}},\ }\href {\doibase 10.1103/physreva.70.012706} {\bibfield
  {journal} {\bibinfo  {journal} {Phys. Rev. A}\ }\textbf {\bibinfo {volume}
  {70}},\ \bibinfo {pages} {012706} (\bibinfo {year} {2004})}\BibitemShut
  {NoStop}%
\bibitem [{\citenamefont {Han}\ \emph {et~al.}(2008)\citenamefont {Han},
  \citenamefont {Zhong}, \citenamefont {Zhang},\ and\ \citenamefont
  {Shi}}]{han2008ss}%
  \BibitemOpen
  \bibfield  {author} {\bibinfo {author} {\bibfnamefont {H.}~\bibnamefont
  {Han}}, \bibinfo {author} {\bibfnamefont {Z.}~\bibnamefont {Zhong}}, \bibinfo
  {author} {\bibfnamefont {X.}~\bibnamefont {Zhang}}, \ and\ \bibinfo {author}
  {\bibfnamefont {T.}~\bibnamefont {Shi}},\ }\href {\doibase
  10.1103/physreva.78.044701} {\bibfield  {journal} {\bibinfo  {journal} {Phys.
  Rev. A}\ }\textbf {\bibinfo {volume} {78}},\ \bibinfo {pages} {044701}
  (\bibinfo {year} {2008})}\BibitemShut {NoStop}%
\bibitem [{\citenamefont {Hu}\ \emph {et~al.}(2016)\citenamefont {Hu},
  \citenamefont {Yao}, \citenamefont {Wang}, \citenamefont {Li}, \citenamefont
  {Gu},\ and\ \citenamefont {Zhong}}]{hu2016variational}%
  \BibitemOpen
  \bibfield  {author} {\bibinfo {author} {\bibfnamefont {M.-H.}\ \bibnamefont
  {Hu}}, \bibinfo {author} {\bibfnamefont {S.-M.}\ \bibnamefont {Yao}},
  \bibinfo {author} {\bibfnamefont {Y.}~\bibnamefont {Wang}}, \bibinfo {author}
  {\bibfnamefont {W.}~\bibnamefont {Li}}, \bibinfo {author} {\bibfnamefont
  {Y.-Y.}\ \bibnamefont {Gu}}, \ and\ \bibinfo {author} {\bibfnamefont {Z.-X.}\
  \bibnamefont {Zhong}},\ }\href {\doibase 10.1016/j.cplett.2016.05.017}
  {\bibfield  {journal} {\bibinfo  {journal} {Chem. Phys. Lett.}\ }\textbf
  {\bibinfo {volume} {654}},\ \bibinfo {pages} {114} (\bibinfo {year}
  {2016})}\BibitemShut {NoStop}%
\bibitem [{\citenamefont {Doolen}(1975)}]{doolen1975procedure}%
  \BibitemOpen
  \bibfield  {author} {\bibinfo {author} {\bibfnamefont {G.~D.}\ \bibnamefont
  {Doolen}},\ }\href {\doibase 10.1088/0022-3700/8/4/010} {\bibfield  {journal}
  {\bibinfo  {journal} {J. Phys. B}\ }\textbf {\bibinfo {volume} {8}},\
  \bibinfo {pages} {525} (\bibinfo {year} {1975})}\BibitemShut {NoStop}%
\bibitem [{\citenamefont {B\"urgers}\ and\ \citenamefont
  {Lindroth}(2000)}]{burgers2000doubly}%
  \BibitemOpen
  \bibfield  {author} {\bibinfo {author} {\bibfnamefont {A.}~\bibnamefont
  {B\"urgers}}\ and\ \bibinfo {author} {\bibfnamefont {E.}~\bibnamefont
  {Lindroth}},\ }\href {\doibase 10.1007/s100530050556} {\bibfield  {journal}
  {\bibinfo  {journal} {Eur. Phys. J. D}\ }\textbf {\bibinfo {volume} {10}},\
  \bibinfo {pages} {327} (\bibinfo {year} {2000})}\BibitemShut {NoStop}%
\bibitem [{\citenamefont {Li}\ and\ \citenamefont
  {Shakeshaft}(2005)}]{li2005s}%
  \BibitemOpen
  \bibfield  {author} {\bibinfo {author} {\bibfnamefont {T.}~\bibnamefont
  {Li}}\ and\ \bibinfo {author} {\bibfnamefont {R.}~\bibnamefont
  {Shakeshaft}},\ }\href {\doibase 10.1103/physreva.71.052505} {\bibfield
  {journal} {\bibinfo  {journal} {Phys. Rev. A}\ }\textbf {\bibinfo {volume}
  {71}},\ \bibinfo {pages} {052505} (\bibinfo {year} {2005})}\BibitemShut
  {NoStop}%
\bibitem [{\citenamefont {Kar}\ and\ \citenamefont {Ho}(2012)}]{kar2012shape}%
  \BibitemOpen
  \bibfield  {author} {\bibinfo {author} {\bibfnamefont {S.}~\bibnamefont
  {Kar}}\ and\ \bibinfo {author} {\bibfnamefont {Y.~K.}\ \bibnamefont {Ho}},\
  }\href {\doibase 10.1103/physreva.86.014501} {\bibfield  {journal} {\bibinfo
  {journal} {Phys. Rev. A}\ }\textbf {\bibinfo {volume} {86}},\ \bibinfo
  {pages} {014501} (\bibinfo {year} {2012})}\BibitemShut {NoStop}%
\bibitem [{\citenamefont {Ho}(1979)}]{ho1979autoionization}%
  \BibitemOpen
  \bibfield  {author} {\bibinfo {author} {\bibfnamefont {Y.~K.}\ \bibnamefont
  {Ho}},\ }\href {\doibase 10.1103/physreva.19.2347} {\bibfield  {journal}
  {\bibinfo  {journal} {Phys. Rev. A}\ }\textbf {\bibinfo {volume} {19}},\
  \bibinfo {pages} {2347} (\bibinfo {year} {1979})}\BibitemShut {NoStop}%
\bibitem [{\citenamefont {Ho}(1981)}]{ho1981complex}%
  \BibitemOpen
  \bibfield  {author} {\bibinfo {author} {\bibfnamefont {Y.~K.}\ \bibnamefont
  {Ho}},\ }\href {\doibase 10.1103/physreva.23.2137} {\bibfield  {journal}
  {\bibinfo  {journal} {Phys. Rev. A}\ }\textbf {\bibinfo {volume} {23}},\
  \bibinfo {pages} {2137} (\bibinfo {year} {1981})}\BibitemShut {NoStop}%
\bibitem [{\citenamefont {Gning}\ \emph {et~al.}(2015)\citenamefont {Gning},
  \citenamefont {Sow}, \citenamefont {Traor{\'{e}}}, \citenamefont {Dieng},
  \citenamefont {Diakhate}, \citenamefont {Biaye},\ and\ \citenamefont
  {Wagu{\'{e}}}}]{GNING20151}%
  \BibitemOpen
  \bibfield  {author} {\bibinfo {author} {\bibfnamefont {Y.}~\bibnamefont
  {Gning}}, \bibinfo {author} {\bibfnamefont {M.}~\bibnamefont {Sow}}, \bibinfo
  {author} {\bibfnamefont {A.}~\bibnamefont {Traor{\'{e}}}}, \bibinfo {author}
  {\bibfnamefont {M.}~\bibnamefont {Dieng}}, \bibinfo {author} {\bibfnamefont
  {B.}~\bibnamefont {Diakhate}}, \bibinfo {author} {\bibfnamefont
  {M.}~\bibnamefont {Biaye}}, \ and\ \bibinfo {author} {\bibfnamefont
  {A.}~\bibnamefont {Wagu{\'{e}}}},\ }\href {\doibase
  10.1016/j.radphyschem.2014.06.015} {\bibfield  {journal} {\bibinfo  {journal}
  {Radiat. Phys. Chem.}\ }\textbf {\bibinfo {volume} {106}},\ \bibinfo {pages}
  {1} (\bibinfo {year} {2015})}\BibitemShut {NoStop}%
\bibitem [{\citenamefont {Ho}\ and\ \citenamefont
  {Bhatia}(1991)}]{ho1991complex}%
  \BibitemOpen
  \bibfield  {author} {\bibinfo {author} {\bibfnamefont {Y.~K.}\ \bibnamefont
  {Ho}}\ and\ \bibinfo {author} {\bibfnamefont {A.~K.}\ \bibnamefont
  {Bhatia}},\ }\href {\doibase 10.1103/physreva.44.2895} {\bibfield  {journal}
  {\bibinfo  {journal} {Phys. Rev. A}\ }\textbf {\bibinfo {volume} {44}},\
  \bibinfo {pages} {2895} (\bibinfo {year} {1991})}\BibitemShut {NoStop}%
\bibitem [{\citenamefont {Drake}\ and\ \citenamefont
  {Makowski}(1988)}]{Drake1988}%
  \BibitemOpen
  \bibfield  {author} {\bibinfo {author} {\bibfnamefont {G.~W.~F.}\
  \bibnamefont {Drake}}\ and\ \bibinfo {author} {\bibfnamefont {A.~J.}\
  \bibnamefont {Makowski}},\ }\href {\doibase 10.1364/josab.5.002207}
  {\bibfield  {journal} {\bibinfo  {journal} {J. Opt. Soc. Am. B}\ }\textbf
  {\bibinfo {volume} {5}},\ \bibinfo {pages} {2207} (\bibinfo {year}
  {1988})}\BibitemShut {NoStop}%
\bibitem [{\citenamefont {Yan}\ \emph {et~al.}(2003)\citenamefont {Yan},
  \citenamefont {Zhang},\ and\ \citenamefont {Li}}]{yan2003energies}%
  \BibitemOpen
  \bibfield  {author} {\bibinfo {author} {\bibfnamefont {Z.-C.}\ \bibnamefont
  {Yan}}, \bibinfo {author} {\bibfnamefont {J.-Y.}\ \bibnamefont {Zhang}}, \
  and\ \bibinfo {author} {\bibfnamefont {Y.}~\bibnamefont {Li}},\ }\href
  {\doibase 10.1103/physreva.67.062504} {\bibfield  {journal} {\bibinfo
  {journal} {Phys. Rev. A}\ }\textbf {\bibinfo {volume} {67}},\ \bibinfo
  {pages} {062504} (\bibinfo {year} {2003})}\BibitemShut {NoStop}%
\bibitem [{\citenamefont {Fletcher}(2000)}]{fletcher2013practical}%
  \BibitemOpen
  \bibfield  {author} {\bibinfo {author} {\bibfnamefont {R.}~\bibnamefont
  {Fletcher}},\ }\href {\doibase 10.1002/9781118723203} {\emph {\bibinfo
  {title} {Practical Methods of Optimization}}}\ (\bibinfo  {publisher} {John
  Wiley {\&} Sons, Ltd},\ \bibinfo {year} {2000})\BibitemShut {NoStop}%
\end{thebibliography}%

\end{document}